\documentstyle[12pt]{article}
\newcommand{\nn}{\nonumber \\}
\newcommand{\R}{{\bf R}}
\newcommand{\C}{{\bf C}}
\newcommand{\Z}{{\bf Z}}
\newcommand{\ar}{\left(\begin{array}{c}}
\newcommand{\arr}{\left(\begin{array}{cc}}
\newcommand{\arrr}{\left(\begin{array}{ccc}}
\newcommand{\ay}{\end{array}\right)}

\newcommand{\point}{$ \! \! \! \! \! \! \! \! \! $ . $\,$}
\newcommand{\be}{\begin{equation}}
\def\l#1{\label{eq:#1}\end{equation}}
\def\r#1{(\ref{eq:#1})}
\def\lbq#1{\label{eq:#1}}
\begin{document}

\title
{\bf Relativistic Quantum Mechanics on the SL(2, R) Spacetime} 
\author{T. F\"ul\"op\thanks{E-mail: fulopt@hal9000.elte.hu}
      \\ {\it Institute for Theoretical Physics} \\
         {\it E\"otv\"os University, Budapest, Hungary}}
\date{September 10, 1996}
\maketitle

    \begin{abstract}
 
The Schr\"odinger-type formalism of the Klein-Gordon quantum mechanics is 
adapted for the case of the $SL(2,\R)$ spacetime. The free particle case is 
solved, the results of a recent work are reproduced while all the other, 
topologically nontrivial solutions and the antiparticle modes are also 
found, and a deeper insight into the physical content of the theory is 
given.

    \end{abstract}

    \section{\point Introduction}\label{intro}

Recently the classical and quantum mechanics of the zero spin particle
moving freely on the $SL(2,\R)$ group manifold were examined in
\cite{JOT}. The theory of the system was constructed via Hamiltonian
reduction, a method becoming increasingly popular nowadays in the field of
$\cal W$-algebras and integrable models (for references see \cite{JOT}).
This problem is of interest because of a number of aspects. {}From the
point of view of conformal field theory this system is the point particle
analogue of the $SL(2,\R)$ WZNW model, having an analogous
reparametrization invariance property. {}From the aspect of string theory
the system can be considered to describe the 'centre-of-mass motion' of a
(WZNW) string on the $SL(2,\R)$ spacetime. The problem is also of interest
from the point of view of general relativity because it offers an example
of an exactly solvable quantum system on a curved spacetime background. 

After constructing the classical theory, the authors in \cite{JOT} quantize 
the system by considering unitary, irreducible representations of the 
algebra formed by the symmetry currents corresponding to the left and right 
translation invariance. The results set some interesting problems and 
questions. One of them is that, due to the representations given in 
\cite{JOT}, the value of the mass of the particle cannot be arbitrary. Only 
a discrete series is allowed for the possible mass values. It is plausible 
to conjecture that this condition is of topological origin. The 
topology of the $SL(2,\R)$ group manifold is $ \R^2 \times {\cal S}^1 $, 
and the presence of a compact dimension would be responsible for the mass 
quantization condition. Then we may ask whether this condition is 
necessary for a consistent solution or the other mass values also 
correspond to additional - consistent, while topologically 'nontrivial' - 
solutions. The idea is that if the condition holds then the wave function 
of the particle is single-valued around the ${\cal S}^1$ direction, while 
for other mass values a nontrivial constant phase factor would be present.

Another question is that one expects the appearance of antiparticle modes,
similarly as in the case of Minkowski spacetime. It would be nice to find
them, too, as a natural part of the complete space of states. Thirdly, in
the usual cases of nonrelativistic quantum mechanics or quantum mechanics
on Minkowski spacetime the state space, the Hamiltonian, etc.\ of a quantum
system suppose that initially an inertial reference frame has been
chosen. What is the corresponding step in the case of the $SL(2,\R)$
spacetime? Special attention has to be payed to this problem as the
spacetime in question is a curved one. 

Finally, it would be interesting to answer some other questions concerning 
the formalism, including how to define observables and expectation values
or how to handle the case when an external field is present. 

In this paper these questions will be answered. Our method is the
adaptation of the Schr\"odinger-type formalism of the special relativistic
Klein-Gordon quantum mechanics to the case of the $SL(2,\R)$ spacetime.
The advantage of using this extension of the Klein-Gordon framework is
that in this way the physical properties of the system become more
transparent. This formalism makes the particle and antiparticle modes and
the corresponding charge symmetry more visible, and provides the theory
a consistent physical interpretation (including, e.g., the 
definition of observables and expectation values). 

The paper is organized as follows: in Sec.\ II the results of \cite{JOT} 
are presented. The necessary ingredients of the formalism of the quantum 
mechanics of the spin zero particle on Minkowski spacetime are summarized 
in  Sec.\ III\@. The properties of the $SL(2,\R)$ spacetime and the choice 
of a suitable reference frame on it are discussed in Sec.\ IV\@. In Sec.\ 
V the formalism presented in Sec.\ III is set on the $SL(2,\R)$ spacetime. 
The resulting theory is solved in the free particle case in Sec.\ VI\@. 
The left and right symmetric aspects of the free system are discussed in 
Sec.\ VII\@. 

    \section{\point Quantization via Hamiltonian reduction}\label{JOT}

$SL(2,\R)$ is a three dimensional Lie group, the naturally arising metric 
is 
    \be 
    h_{\mu\nu}(x) = \frac{1}{2} \mbox{Tr} \left[ g^{-1} \partial_\mu g 
    g^{-1} \partial_\nu g \right]
    \l{metric} 
(with a local parametrization $ x \mapsto g(x) \in SL(2,\R) $). This smooth 
metric is nondegenerate and proves to be of Lorentz signature, thus 
$SL(2,\R)$ can be considered as a $ 2 + 1 $ dimensional curved spacetime. 

The classical and the quantum theory of a free particle on the $SL(2,\R)$
spacetime is constructed in \cite{JOT} as follows. Classically the action
    \be
    I_0 = - \kappa \int dt \sqrt { h_{\mu\nu} (x) \dot x^\mu \dot x^\nu }
    \l{hatas}
describes a particle of mass $ \kappa > 0 $, with $ x^\mu (t) $ denoting the
trajectory of the particle. The action $I_0$ can be obtained from a more 
appropriate quadratic action $I$ by imposing a given constraint. This first 
class constraint arises as a consistency condition on the canonical momenta 
of the system $I$, and the local gauge symmetry it generates is nothing but 
the reparametrization invariance of the system. Then, according to the 
method of Hamiltonian reduction, the reduced phase space can be obtained by 
factorizing the constrained surface with respect to the gauge symmetry.  

In addition to the reparametrization invariance the system is invariant 
under the left and right transformations $ g \mapsto f g $, $ g \mapsto g 
\tilde f^{-1} $, $ f, \tilde f \in SL(2,\R) $. The reduced phase space is 
found to be of the form $ {\cal O}_K \times {\cal O}_{-K} $ where $ {\cal 
O}_K $ and $ {\cal O}_{-K} $ are the coadjoint orbits of the left, resp.\ 
the right, transformations passing through a fixed timelike vector $K$ 
arbitrarily chosen from $sl(2,\R)$, the Lie algebra of $SL(2,\R)$. The left 
and right symmetry currents parametrize the reduced phase space and form 
two independent $sl(2,\R)$ algebras under the Poisson brackets. 

The quantum theory is obtained in \cite{JOT} via quantizing the reduced
phase space. Unitary, irreducible representations of the algebra
$sl(2,\R)$ formed by the symmetry currents on ${\cal O}_K$ and ${\cal
O}_{-K}$ are sought. The state space is spanned by the tensorial product
of two such representations, which have to share the same value for the
Casimir due to a mass shell condition. Such a state (denoted by $ \vert
n_L, n_R \rangle $) is labelled by two non-negative integers. The quantum
commutation relations for the left current in an appropriate basis are
    \be
    [ L_0 , L_1 ] = -2i L_2, \qquad
    [ L_1 , L_2 ] = 2i L_0, \qquad
    [ L_2 , L_0 ] = -2i L_1.
    \l{leftcomm}
With $ L_{\pm} = L_1 \mp i L_2 $ one finds
    \begin{eqnarray}\lbq{balhatas}
    L_0 \vert n_L, n_R \rangle & = & 2 (n_L + j) \vert n_L, n_R \rangle,
    \nn
    L_+ \vert n_L, n_R \rangle & = & 2 \sqrt{ (n_L + 2j) (n_L + 1) } 
    \vert n_L + 1 , n_R \rangle, \\
    L_- \vert n_L, n_R \rangle & = & 2 \sqrt{ (n_L - 1 + 2j) n_L } 
    \vert n_L - 1 , n_R \rangle. \nonumber
    \end{eqnarray}
In a similar manner, the action of the right current on the states is
    \begin{eqnarray}
    R_0 \vert n_L, n_R \rangle & = & - 2 (n_R + j) \vert n_L, n_R \rangle,
    \nn
    R_- \vert n_L, n_R \rangle & = & - 2 \sqrt{ (n_R + 2j) (n_R + 1) } 
    \vert n_L, n_R + 1 \rangle, \\
    R_+ \vert n_L, n_R \rangle & = & - 2 \sqrt{ (n_R - 1 + 2j) n_R } 
    \vert n_L, n_R - 1 \rangle, \nonumber
    \lbq{jobbhatas}\end{eqnarray}
where $ j = \frac{1}{2} ( 1 + \sqrt{ 1 + \kappa^2 } ) $. As in \cite{JOT} 
the left and right representations are chosen from the discrete series $ 
D_j^{\pm} $ \cite{Ber}, $j$ has to take one of the values $ \frac{3}{2}, 
\frac{4}{2}, \frac{5}{2}, \ldots $ Consequently, the mass of the particle 
is not arbitrary but must come from a discrete series of allowed values. 

The states $ \vert n_L, n_R \rangle $ are eigenstates of the energy and the 
angular momentum, which operators can be identified as $ \frac{1}{2} ( L_0 
- R_0 ) $ resp.\ $ \frac{1}{2} ( L_0 + R_0 ) $, the corresponding 
eigenvalues are $ n_L + n_R + 2j $ resp.\ $ n_L - n_R $. The energy levels 
are positive definite and spaced integrally, the angular momentum takes 
integer values. 

    \section{\point The quantum mechanics of the zero spin particle on 
Minkowski spacetime}\label{FV}

To recall the basic elements of the Schr\"odinger-type formalism of the 
quantum mechanics of the zero spin particle on Minkowski spacetime we follow 
the approach of Feshbach and Villars \cite{FV}. This formalism is the close 
analogy of the one of the spin $ \frac{1}{2} $ particle case and gives a 
consistent and well-interpretable one-particle quantum theory. 

We start from the Klein-Gordon equation for a particle with mass $\kappa$ 
and electric charge $e$ in the presence of an electromagnetic potential $ 
A^\mu = ( A^0, A^k ) = ( A^0, \mbox{\boldmath $A$} ) $:
    \be\lbq{KG}
    ( D^\mu D_\mu - \kappa^2 ) \psi = 0
    \end{equation}
where $ D_\mu = \frac{\partial}{\partial x^\mu} - i e A_\mu $ (we work in $ 
\hbar = c = 1 $). Our purpose is to reformulate \r{KG} as an equation of 
the form $ i ( \partial \Psi / \partial t ) = H \Psi $ with an appropriate 
$\Psi$. This can be achieved by considering $ D_0 \psi $ as an independent 
degree of freedom and introducing the two-component column vector $\Psi$ 
with components $\psi$ and $ D_0 \psi $. More precisely, later convenience 
suggests to define $\Psi$ as
    \be
    \Psi = \ar \varphi \\ \chi \ay = \frac{1}{\sqrt{2}} \ar \psi +  
    \frac{i}{\kappa} D_0 \psi \\ \psi - \frac{i}{\kappa} D_0 \psi \ay .
    \l{Psi}
Then \r{KG} can be rewritten as 
    \begin{eqnarray}\lbq{eqnsofmotion}
    i ( \partial \varphi / \partial t ) & = & \frac{1}{2\kappa}
    \left( \frac{1}{i} \mbox{ \boldmath $\nabla$ } - e \mbox{\boldmath 
    $A$} \right)^2 (
    \varphi + \chi ) + ( \kappa - e A_0 ) \varphi, \nn i ( \partial \chi /
    \partial t ) & = & - \frac{1}{2\kappa} \left( \frac{1}{i} \mbox{ \boldmath
    $\nabla$ } - e \mbox{\boldmath $A$} \right)^2 ( \varphi + \chi ) - 
    ( \kappa + e A_0 ) \chi,
    \end{eqnarray}
the operator $H$ can be read off from \r{eqnsofmotion}. 

{}From \r{KG} the current four-vector defined as $ j_\mu = const. ( \psi^* 
D_\mu \psi - \mbox{c.c.} ) $ proves to be conserved. One can express 
$j_\mu$ with $\Psi$ as well, for example the density reads $ j_0 = \Psi^* 
\sigma_3 \Psi = \varphi^* \varphi - \chi^* \chi $ ($\sigma_k$ denotes the 
Pauli matrices). The density $j_0$ is not positive definite, $j_\mu$ is 
interpreted not as the probability current but as the charge current. If 
$\Psi$ satisfies the equation  $ i ( \partial \Psi / \partial t ) = H (e) 
\Psi $ then the charge conjugate wave function 
    \be
    \Psi_c = \ar \chi^* \\ \varphi^* \ay
    \l{Psic}
satisfies $ i ( \partial \Psi_c / \partial t ) = H (-e) \Psi_c $ (where 
$H(-e)$ differs from $H(e)$ only by the sign of the electric charge), the 
density corresponding to $\Psi_c$ is $ \Psi_c^* \sigma_3 \Psi_c = - \Psi^* 
\sigma_3 \Psi $. Thus we can see that the theory actually describes two 
degrees of freedom with opposite charges (a particle and an antiparticle) 
and has a fundamental charge symmetry. The advantage of using $\varphi$ and 
$\chi$ as the two components of the wave function is that this property 
becomes apparent. In the nonrelativistic limit the two degrees of freedom 
decouple and lead to two independent Schr\"odinger 
equations of the usual form. 

If $ \int \Psi^* \sigma_3 \Psi d^3 x $ is positive/negative then let $\Psi$ 
be called 'positive', resp.\ 'negative', and be normalized so that this 
integral be $+1$ resp.\ $-1$, expressing that the charge of such a state is 
$+e$ resp.\ $-e$. 

The inner product that turns out to be appropriate for this formalism is 
    \be
    ( \Psi_1 , \Psi_2 ) = \int \Psi_1^* \sigma_3 \Psi_2^{\,} d^3 x = 
    \int \left( \varphi_1^* \varphi_2^{\,} - \chi_1^* \chi_2^{\,} 
    \right) d^3 x.
    \l{(,)}
This inner product is not positive definite, the space of the wave 
functions is not a Hilbert space, on the contrary to the nonrelativistic 
case. Physical quantities correspond to Hermitian operators acting on the 
$\Psi$-s where Hermiticity, the expectation value of an operator and such 
notions are defined with respect to the inner product \r{(,)}. An 
interesting consequence of the indefiniteness of the inner product is that 
the eigenvalues of an operator are not necessarily expectation values as 
well. For example, if $ A_\mu = 0 $ then the expectation value of the 
Hamiltonian in an eigenstate with eigenvalue $E$ is $ \vert E \vert > 0 $. 
Similarly, the expectation value of the kinetic energy operator of the 
general case $ A_\mu \neq 0 $ always proves to be positive definite. 

We remark that further interpretation issues and the case of the neutral 
particles are also discussed in \cite{FV}. 

    \section{\point The SL(2, R) spacetime}

The Schr\"odinger-type formalism of the Klein-Gordon quantum mechanics 
required the choice of an inertial reference frame on the Minkowski 
spacetime. As we want to adapt this formalism for the 
$SL(2,\R)$ spacetime, we have to solve the nontrivial problem of finding 
an analogous step on the curved spacetime of $SL(2,\R)$. 

Let us consider the choice of a reference frame on the Minkowski 
spacetime the following way. We start by choosing a timelike vector field 
on the spacetime. Then we construct spacelike hypersurfaces being 
orthogonal to the integral curves corresponding to this vector field. The 
time coordinate is introduced as the parameter along the integral curves 
and the space coordinates are introduced to parametrize the hypersurfaces. 
An inertial reference frame is such a special reference frame that the 
vector field chosen is constant, the same timelike vector (absolute or 
four-velocity) is attached to each spacetime point, the integral curves are 
parallel straight lines and the orthogonal hypersurfaces are parallel 
hyperplanes. In other words, we produce an inertial reference frame by 
considering an absolute velocity vector in a spacetime point and then we 
shift this timelike vector by a parallel translation to all the other 
spacetime points. 

It is this translation idea that is appropriate to define the analogue of
an inertial reference frame on $SL(2,\R)$. Let us choose its identity
element as one of the spacetime points and let us consider a timelike
vector $U$ in its tangent space, i.e., in $sl(2,\R)$. Then we shift this
vector to the tangent space of a spacetime point $g$ by the natural
translation $gU$ (where this multiplication means simply the
multiplication of matrices). However, an important difference from the
Minkowski case is that now the left and the right translation are not
equal, $ gU \neq Ug $, as a consequence of the noncommutativity of the
group $SL(2,\R)$. Thus we can define a "left-inertial reference frame" and
a "right-inertial reference frame" corresponding to the two possibilities
of translation. Moreover, not only $gU$ and $Ug$ can be chosen naturally,
but also the average $ \frac{1}{2} (gU + Ug) $ (the "middle-inertial
reference frame").  Straightforward calculations show that for each of
these three timelike Killing vector fields the metric \r{metric} is
time-independent. The middle-inertial reference frame has an additional
advantageous property:  the space-time mixed components of $h_{\mu\nu}$
prove to be zero. That is why we choose the middle-inertial reference
frame for the following considerations. 

Now let us make use of the fact that, for a fixed $U$, a $ g \in SL(2,\R) $ 
can be given in the form $ e^{ \frac{1}{2} t U } e^C e^{ \frac{1}{2} t U } 
$ where $ t \in [ 0, 2\pi ) $ and $ C \in sl(2,\R) $, $C$ is orthogonal to 
$U$; any such $t$ and $C$ uniquely characterizes an element of $SL(2,\R)$. 
(This statement can be proven with the aid of the formula $ e^{ \xi^k 
\sigma_k} = \cosh R \: {\bf 1} + \sinh R /R \; \xi^k \sigma_k $, $ R^2 = 
\xi^k \xi^k $.) Accordingly, $ \partial_t g = \frac{1}{2} ( gU + Ug ) $ 
($C$ is kept fixed). Thus by a coordinatization $ C = C ( x^1, x^2 ) $ and 
with $ x^0 := t $ we obtain a middle-inertial coordinate system 
corresponding to $U$. For example, in polar coordinates: $ x^1 = r $, $ x^2 
= \vartheta $, $ C = r \cos \vartheta X + r \sin \vartheta Y $ [with $X$ 
and $Y$ fixed such that $ (U, X, Y) $ is an orthonormal basis in 
$sl(2,\R)$] the metric tensor reads
    \be
    \{ h_{\mu\nu} \} = \arrr
    -\cosh^2 r & 0 & 0 \\ 0 & 1 & 0 \\ 0 & 0 & \sinh^2 r \ay .
    \l{polarmetric}
    
Remarkably, the topology of $SL(2,\R)$ is $ \R^2 \times {\cal S}^1 $---the 
timelike geodesics are the closed ones---, hence this manifold cannot be 
covered with a single open coordinate patch. However, a middle-inertial 
coordinate system covers the whole $SL(2,\R)$ if we identify $ t = 2 \pi $ 
with $ t = 0 $. This way we can avoid the use of multiple patches. 

We mention that the elements of $SL(2,\R)$ can be given also in the form $
e^{ \xi X } e^{ \eta Y } e^{ \tau U } $, where $ \xi, \eta \in \R $ and $
\tau \in [ 0, 2\pi ) $. Later we will make use of this fact, too. 
    
    \section{\point Relativistic quantum mechanics on the SL(2, R) 
spacetime}

To build up the relativistic quantum theory on the $SL(2,\R)$ spacetime we 
follow the steps of Sec.\ III\@. Now the Klein-Gordon equation reads
    \be
    ( \tilde D^\mu \tilde D_\mu - \kappa^2 ) \psi = 0
    \l{SLKG}
where $ \tilde D_\mu = \nabla_\mu - i e A_\mu $ ($ \nabla_\mu $ denotes the 
covariant derivative). In a middle-inertial coordinate system the metric is 
time independent and its space-time mixed components are zero, thus 
re-writing \r{SLKG} as a first order equation in the variable $\Psi$ 
introduced as in the flat case (cf. \r{Psi}, $ D_\mu = \partial_\mu - i 
e A_\mu $) yields
    \begin{eqnarray}
    i ( \partial \varphi / \partial t ) = \frac{1}{2\kappa} \frac{1}{h^{00}
    \sqrt{-h}} D_j \left( \sqrt{-h} h^{jk} D_k \right) ( \varphi + 
    \chi ) \nn
    + \frac{\kappa}{2} \left[ \left( 1 - \frac{1}{h^{00}} \right) \varphi -
    \left( 1 
    + \frac{1}{h^{00}} \right) \chi \right] - e A_0 \varphi, \nn 
    i ( \partial 
    \chi / \partial t ) = - \frac{1}{2\kappa} \frac{1}{h^{00}
    \sqrt{-h}} D_j \left( \sqrt{-h} h^{jk} D_k \right) ( \varphi + 
    \chi ) \nn
    + \frac{\kappa}{2} \left[ \left( 1 + \frac{1}{h^{00}} \right) \varphi +
\left( 
    \frac{1}{h^{00}} - 1 \right) \chi \right] - e A_0 \chi 
    \lbq{SLeqnsofmotion}\end{eqnarray}
where $h$ is the determinant of the metric, $ j, k = 1, 2 $. The 
corresponding Hamiltonian can be read off from \r{SLeqnsofmotion}. 

The inner product to be introduced as the analogue of the special 
relativistic one must satisfy some requirements, i.e., to be invariant under 
space $ \rightarrow $ space transformations $ ( x^1 , x^2 ) \rightarrow ( 
{x^1}' , {x^2}' ) $ and to ensure that the Hamiltonian be symmetric and the 
charge be conserved. The result is the appearance of a weight function in 
the integral \r{(,)}:
    \be
    ( \Psi_1 , \Psi_2 ) = \int \Psi_1^* \sigma_3 \Psi_2^{\,} \sqrt{-h}
    \left( -h^{00} \right) d^2 x = \int \left( \varphi_1^* \varphi_2^{\,}
    - \chi_1^* \chi_2^{\,} \right) \sqrt{-h} \left( -h^{00} \right) d^2 x. 
    \l{(,)'}
   
It can be checked easily that the fundamental charge symmetry keeps valid 
without any modifications of the formulae of the Minkowski case. Similarly, 
the definition of observables and expectation values, as well as all other 
interpretation issues, also can be adapted appropriately from the special 
relativistic formalism. 

We mention that the considerations of this section are applicable not only 
for the $SL(2,\R)$ spacetime but, more generally, for any static spacetime, 
i.e., for such curved spacetimes that have a time independent metric with 
zero space-time mixed components (in a suitable coordinatization). 

    \section{\point The free masspoint}
    
Now let us solve the eigenvalue problem for the Hamiltonian of the free 
system. The free Hamiltonian is time independent---following from the time 
independence of the metric---, thus $ H \Psi = E \Psi $ implies $ \Psi(t) = 
\exp(-iEt) \Psi(0) $ (in the following this time dependence will always be 
understood to the eigenfunctions). Thus $ \varphi(t) = \exp(-iEt) 
\varphi(0) $ and 
$ \chi(t) = \exp(-iEt) \chi(0) $, and, consequently, $ \psi(t) = \exp(-iEt) 
\psi(0) $. As a result, an eigenfunction $\Psi$ can be expressed by means 
of $\psi$ only (!), from \r{Psi} one finds
    \be \Psi = \frac{1}{\sqrt{2}} \ar 1 + \frac{E}{\kappa} \\ 1 -
    \frac{E}{\kappa} \ay \psi = \ar \frac{1}{\sqrt{2}} \left( 1 +
    \frac{E}{\kappa} \right) \psi \\ \frac{1}{\sqrt{2}} \left( 1 - 
    \frac{E}{\kappa} \right) \psi \ay. 
    \l{eigenPsi}
The inner product of a $\Psi_1$ with eigenvalue $E_1$ and a $\Psi_2$ with 
eigenvalue $E_2$ can also be expressed with the corresponding $\psi_1$ and 
$\psi_2$:  
    \be
    ( \Psi_1 , \Psi_2 ) = \frac{ E_1 + E_2 }{\kappa} \int \psi_1^* \psi_2^{\,}
    \sqrt{-h} \left( -h^{00} \right) d^2 x.
    \l{PsiPsipsipsi}
Substituting the connection between $\Psi$ and $\psi$ into $ H \Psi = E 
\Psi $ and working in polar coordinates - an 
appropriate concrete space coordinatization, cf. \r{polarmetric} - gives
    \be
    \partial_r^2 \psi + 2 \coth 2r \partial_r \psi + 
    \frac{1}{\sinh^2 r} \partial^2_\vartheta
    \psi + \frac{E^2}{\cosh^2 r} \psi - \kappa^2 \psi = 0.
    \l{eigenpsi}
    
We can expand $\psi$ into Fourier series in the variable $ \vartheta \in 
[0,2\pi) $, thus expressing it as a linear combination of the functions $ 
\exp (i m \vartheta) $, $ m \in \Z $. A term of this series is of the form 
    \be 
    \psi_{m,E} = W_{m,E} (r) e^{ i m \vartheta } e^{ - i E t }, 
    \l{common}
the corresponding $\Psi_{m,E}$ is an eigenfunction of the angular momentum 
operator $ J = -i \partial_\vartheta $ as well, hence determining all the 
$\Psi_{m,E}$-s means the common diagonalization of $H$ and $J$. 

Eq. \r{eigenpsi} implies an ordinary differential equation on $W_{m,E}$. 
This equation can be turned into the hipergeometric equation $ z (1-z) d^2 
w / dz^2 + [ c - ( a + b + 1 ) z ] dw / dz - ab w = 0 $ in the new variable 
$ z = \tanh^2 r $ and with $ a = j + ( \vert E \vert + \vert m \vert ) / 2 
$, $ b = j + ( - \vert E \vert + \vert m \vert ) / 2 $, $ c = 1 + \vert m 
\vert $, $ w = z^{(1-c)/2} (1-z)^{(c-a-b-1)/2} W_{m,E} $ where $ j = 
\frac{1}{2} ( 1 + \sqrt{1 + \kappa^2} ) $ (for the conventions and 
properties used concerning the hipergeometric equation see \cite{Abr} or 
\cite{Erd}). A solution of the hipergeometric equation (with fixed $a$, $b$ 
and $c$) is the hipergeometric function $ F(z) \equiv F(a,b;c;z) $ 
\cite{Abr}. We choose 
    \be
    G(z) \equiv G(a,b;c;z) =F(z) \int_1^z 
    \frac{d \zeta}{ \zeta^c (1-\zeta)^{a+b-c+1} F(\zeta)^2 } 
    \l{G}
as a linearly independent solution \cite{Zal}, the other solutions are 
linear 
combinations of $F$ and $G$. $F$ and $G$ are regular on $(0,1)$, in spite 
of eventual nodes of $F$. 

Now let us search for a maximal orthogonal system of the eigenfunctions. 
$\Psi_{m_1, E_1}$ and $\Psi_{m_2, E_2}$ are orthogonal if $ m_1 \neq m_2 $ 
because of their $\vartheta$-dependence. Thus it is enough to examine the 
case $ m_1 = m_2 = m $. Then $ ( \Psi_{m, E_1} , \Psi_{m,E_2} ) $, 
expressing with the corresponding $w_1$ and $w_2$, is
    \be
    const. \int_0^1 z^{c-1} (1-z)^{ a_1 + b_1 - c } w_1^* w_2^{\,} dz
    \l{inprodw}
(now $ a_1 + b_1 = a_2 + b_2 $ and $ c_1 = c_2 = c  $). With the aid of the 
hipergeometric equation it is not hard to prove that 
    \be
    \int_p^q z^{c-1} (1-z)^{a+b-c} w_1^* w_2^{\,} dz = \frac{1}{a_2 b_2 - 
a_1 
    b_1} \left[ z^c (1-z)^{a+b-c+1} \left( w_1^* \frac{d w_2}{dz} - 
    \frac{d w_1^*}{dz} w_2^{\,} \right) \right]_p^q.
    \l{NL}
{}From the power series $ F(z) = 1 + \frac{ab}{c} z + 
\frac{a(a+1)b(b+1)}{2c(c+1)} z^2 + {\cal O} (z^3) $ the $ z \approx 0 $ 
asymptotic behaviour of $F$, $F'$, $G$ and $G'$ can be determined. 
Concerning the $ z \approx 1 $ behaviour of these functions, using 
\cite{Abr} 
one finds $ F(z) \approx k_1 f_1 (z) $, $ F'(z) \approx k_1^{\,} f_1'(z) 
$, $ 
G(z) \approx k_2 f_2 (z) $ and $ G'(z) \approx k_2^{\,} f_2'(z) $ if $ b 
\not\in 
\Z^-_0 $ while $ F(z) \approx k_3 f_2 (z) $, $ F'(z) \approx k_3^{\,} 
f_2'(z) $, 
$ G(z) \approx k_4 f_1 (z) $ and $ G'(z) \approx k_4^{\,} f_1'(z) $ if $ b 
\in \Z^-_0 $. Here 
    \[
    f_1 (z) = 1 - \frac{(c-a)(c-b)}{a+b-c-1} ( 1 - z ), \qquad 
    f_2 (z) = 1 + \frac{ab}{a+b-c+1} ( 1 - z ), 
    \]
    \be
    k_1 = \Gamma(c) \Gamma(a+b-c) / \Gamma(a) \Gamma(b), \qquad k_2 = - 
    [ k_1 (a+b-c) ]^{-1},
    \l{f,k}
    \[
    k_3 = (-1)^{ \vert b \vert } \Gamma(c) \Gamma(a-c+1) / \Gamma(c-b) 
    \Gamma(a+b-c+1), \qquad k_4 = [ k_3 (a+b-c) ]^{-1}
    \]
(we remark that in our case $a$ and $c$ are always positive).

By using these asymptotics we find that for $ p \rightarrow 0 $ \r{NL} 
diverges if at least one of the corresponding $w$-s is a $G$ (except if $ m 
= 0 $ and the other $w$ is an $F$, however, this case proves be of no 
interest) and converges if both $w$-s are $F$-s. Thus only the $F$-s are 
present in an orthogonal system of eigenfunctions. For $ q \rightarrow 1 $ 
the integral \r{NL} of an $F_1$ and an $F_2$ tends to zero if $ b_1, b_2 
\in \Z^-_0 $, diverges if $ b_1, b_2 \not\in \Z^-_0 $, otherwise it tends 
to a nonzero finite value. Consequently, the parameter $b$ of an $F$ 
appearing in an orthogonal system must be a nonpositive integer. To have a 
maximal set of orthogonal eigenfunctions all such $b$-s have to be 
considered, which means $ \vert E \vert = 2j + \vert m \vert $, $2j + \vert 
m \vert + 1$, \ldots Hence for any mass value $ \kappa > 0 $ there exists a
unique maximal orthogonal set of eigenfunctions, namely $ \{ \Psi_{m,E} \: 
\vert \: m = 0, \pm 1, \pm 2, \ldots; E = \pm ( 2j + \vert m \vert ), \pm ( 
2j + \vert m \vert + 1 ), \ldots \} $. 

We mention here that if $b$ is a nonpositive integer then $F(a,b;c;z)$ is 
nothing but the Jacobi polynomial, more precisely, $ F(a,b;c;z) = ( \vert b 
\vert ) ! / (c)_{ \vert b \vert } \, P_{ \vert b \vert }^{ ( c - 1 , a + b 
- c ) } ( 1 - 2z ) $ \cite{Abr}. Two further remarks are that $w_{m,E}$ [$ 
= F(a,b;c;z) $] and the corresponding $W_{m,E}$ does not depend on the sign 
of $m$ and $E$, and that the vanishing of the special case $ ( \Psi_{m,E} , 
\Psi_{m,-E} ) $ is ensured not by the vanishing of the integral in 
\r{inprodw} but by the vanishing of the constant standing before this 
integral [cf. the factor $ ( E_1 + E_2 ) $ in \r{PsiPsipsipsi}]. 

With the aid of \cite{Abr}, \cite{Erd} and the completeness property of the 
Jacobi polynomials one can verify the completeness of this system of 
eigenfunctions.

After normalization the eigenfunctions $ \Psi_{m,E} $ are of the form
    \begin{eqnarray}
    \Psi_{m,E} (t,r,\vartheta) = 
    \frac{ (-i)^{ \vert m \vert } }{\sqrt{2}} \ar 1 + \frac{E}{\kappa} \\
    1 - \frac{E}{\kappa} \ay \sqrt{ \frac{\kappa}{2\pi} \frac{ \Gamma(a)
    \Gamma(c-b) }{ \Gamma(c)^2 \Gamma(a-c+1) \Gamma(1-b) } } \times \nn
    \mbox{} z^{ 
    \frac{c-1}{2} } ( 1 - z )^{ \frac{a+b-c+1}{2} } F(a,b;c;z) e^{ i m 
    \vartheta } e^{ - i E t } 
    \lbq{Psirthetat}\end{eqnarray}
where $ z = \tanh^2 (r) $---the complex phase factor $ (-i)^{ \vert m
\vert } $ is introduced for later convenience. Concerning our
identification $ (t=2\pi) \equiv (t=0) $ we can observe that if $2j$ is
not an integer value then a---space-, $m$- and $E$-independent, hence
fortunately harmless (see Sec.\ VII)---phase factor appears between
$\Psi_{m,E}(t=2\pi)$ and $\Psi_{m,E}(t=0)$. 

Similarly to the Minkowski case, the energy eigenstates with positive 
eigenvalue prove to be 'positive' (see Sec.\ III) and those with negative 
eigenvalue are 'negative'. The positive wave functions are the particle 
states, and the negative ones are the antiparticle states. Here we can 
see how naturally the antiparticle modes appear in our approach. 

Now let us introduce the notation $ \vert k_L, k_R \rangle $ for the 
eigenstate \r{Psirthetat} where $ k_L = ( E - m ) / 2 $, $ k_R = ( E + m ) 
/ 2 $. The possible values of $k_L$ and $k_R$ are such that $ \vert k_L 
\vert, \vert k_R \vert = j$, $j+1$, $j+2, \ldots $ and for positive 
eigenstates 
both $k_L$ and $k_R$ are positive while for a negative state both are 
negative. With these notations 
    \be
    H \vert k_L, k_R \rangle = ( k_L + k_R ) \vert k_L, k_R \rangle 
    \qquad \mbox{and} \qquad 
    J \vert k_L, k_R \rangle = ( k_R - k_L ) \vert k_L, k_R \rangle .
    \l{HJ}

    \section{\point Symmetry properties}

Comparing our results with \cite{JOT} (see Sec.\ II) we can see that our
investigation reproduces the findings of \cite{JOT}, while it gives
account of the antiparticle states and the topologically nontrivial cases
$ j \not\in \{ \frac{3}{2}, \frac{4}{2}, \ldots \} $ as well. (The quantum
numbers $k_L$, $k_R$ provide a bit more convenient possibility for a
common labelling of the positive and negative eigenstates, this is the
reason why we shifted $n_L$ and $n_R$ to $k_L$ and $k_R$ by an appropriate
$ \pm j $.) Also, the choice of a timelike vector $K$ in \cite{JOT}
corresponds here to a choice of a middle-inertial reference frame based on
an absolute velocity value $U$. What is left is to verify the symmetry
properties of the energy eigenstates in our approach, and to discuss why
in \cite{JOT} only the topologically trivial cases are found. 

The left and right translations $ g \mapsto f g $, $ g \mapsto g { \tilde f 
}^{-1} $ naturally lead to the representations $ [ D_l (f) \psi ] ( g ) = 
\psi ( f^{-1} g ) $, $ [ D_r ( \tilde f ) \psi ] ( g ) = \psi ( g \tilde f 
) $ on a $ \psi : SL(2,\R) \rightarrow \C $. The corresponding 
infinitesimal generators, which give a representation of the Lie algebra 
elements $U$, $X$ and $Y$, are
    \begin{eqnarray}
    l_U & = & - \partial_\vartheta - \partial_t, \nn l_X & = & - \cos ( 
    \vartheta + t ) \partial_r + \coth r \sin ( \vartheta + t ) 
    \partial_\vartheta + \tanh r \sin ( \vartheta + t ) \partial_t, \nn
     l_Y & = & - \sin ( \vartheta + t ) \partial_r - \coth r \cos ( 
    \vartheta + t ) \partial_\vartheta - \tanh r \cos ( \vartheta + t ) 
    \partial_t, \nn \\
    r_U & = & - \partial_\vartheta + \partial_t, \nn r_X & = & \cos ( 
    \vartheta - t ) \partial_r - \coth r \sin ( \vartheta - t ) 
    \partial_\vartheta + \tanh r \sin ( \vartheta - t ) \partial_t, \nn
    r_Y & = & \sin ( \vartheta - t ) \partial_r + \coth r \cos ( 
    \vartheta - t ) \partial_\vartheta - \tanh r \cos ( \vartheta - t ) 
    \partial_t. \nonumber 
    \lbq{lr}\end{eqnarray}
The transformations $D_l(f)$, $D_r( \tilde f)$ are symmetries of the 
system, i.e., they transform a solution of \r{SLKG} to another 
solution of it, as can be verified by means of the infinitesimal 
generators. 

We are interested in the representation of the left and right translations 
on the $\Psi$-s, which can be obtained from $D_l$ and $D_r$ using the 
relation between a $\psi$ and the corresponding $\Psi$ [cf. \r{Psi}]:
    \be
    D_L (f) \Psi = \frac{1}{\sqrt{2}} \ar D_l (f) \psi + 
    \frac{i}{\kappa} \partial_t [ D_l (f) \psi ] \\
    D_l (f) \psi - \frac{i}{\kappa} \partial_t [ D_l (f) \psi ]
    \ay,
    \l{DL}
and the analogous formula for $ D_R ( \tilde f ) $. The action of the 
infinitesimal generators of $D_L$ and $D_R$ on a $\Psi$ can be written as
    \be
    L_k \ar \varphi \\ \chi \ay = \ar l_k \varphi + 
    \frac{i}{2\kappa} [ \partial_t , l_k ] ( \varphi + \chi ) \\ l_k \varphi -
    \frac{i}{2\kappa} [ \partial_t , l_k ] ( \varphi + \chi ) \ay
    \l{L}
($k$ stands for $ U, X, Y $) and similarly for $R_k$. As one can check, the 
$L_k$-s and $R_k$-s satisfy the same commutation relations as the $l_k$-s 
and $r_k$-s. The operators $ L_0 = \frac{1}{i} L_U $, $ L_1 = \frac{1}{i} 
L_X $ and $ L_2 = \frac{1}{i} L_Y $ satisfy \r{leftcomm}. The action of 
$L_0$ and $ L_{\pm} = L_1 \mp i L_2 $ on an energy eigenstate can be 
determined by a straightforward if lengthy calculation involving 
\cite{Abr}, the result is
    \begin{eqnarray}\lbq{balhatas'}
    L_0 \vert k_L, k_R \rangle & = & 2 k_L \vert k_L, k_R \rangle, \nn
    L_+ \vert k_L, k_R \rangle & = & 2 \sqrt{ (k_L + j) (k_L + 1 - j) } 
    \vert k_L + 1 , k_R \rangle, \\
    L_- \vert k_L, k_R \rangle & = & 2 \sqrt{ (k_L - 1 + j) (k_L - j) } 
    \vert k_L - 1 , k_R \rangle. \nonumber
    \end{eqnarray}
With the analogously defined operators $R_0$, $R_1$, $R_2$ and $R_\pm$ one
finds
    \begin{eqnarray}
    R_0 \vert k_L, k_R \rangle & = & - 2 k_R \vert k_L, k_R \rangle, \nn
    R_- \vert k_L, k_R \rangle & = & - 2 \sqrt{ (k_R + j) (k_R + 1 - j) } 
    \vert k_L, k_R + 1 \rangle, \\
    R_+ \vert k_L, k_R \rangle & = & - 2 \sqrt{ (k_R - 1 + j) (k_R - j) } 
    \vert k_L, k_R - 1 \rangle. \nonumber
    \lbq{jobbhatas'}\end{eqnarray}

The linear subspaces spanned by the positive, resp.\ the negative, energy 
eigenstates are invariant subspaces of the left and right transformations. 
Hence both the left and the right representations are a direct sum of two 
irreducible representations, similarly to what happens in the case of 
Minkowski spacetime. For the sake of simplicity, in the following we will 
consider the positive subspace and the left translations only, the parallel 
discussion of the negative subspace and/or for the right translations is 
straightforward. 

If $ j \in \{ \frac{3}{2}, \frac{4}{2}, \ldots \} $ then the eigenvalues
of $L_0$ are integers so the operators $ e^{ i \tau L_0 } = e^{ \tau L_U }
$ give a continous representation of the one-parameter subgroup $ e^{ \tau
U } $ of $SL(2,\R)$. Thus the identification of $ \tau = 0 $ and $ \tau =
2\pi $ does not give any problem. For other values of $j$ $ e^{ \tau L_U }
$ is not continous in $ \tau = 0 \equiv 2\pi $. Hence in these cases the
representation \r{balhatas'} of the algebra $sl(2,\R)$ cannot be
exponentiated to a continous representation of the group $SL(2,\R)$. 
However, the eigenvalues of the operator $ \tilde L_0 = L_0 - \{ 2j \} $
are integers ($ \{ 2j \} $ denotes the fractional part of $2j$). 
Correspondingly, for arbitrary $j$-s, let us introduce the mapping $
\tilde D_L^j $ that maps a group element $ e^{ \xi X } e^{ \eta Y } e^{
\tau U } $ to the operator $ e^{ \, \xi L_X } e^{ \eta L_Y } e^{ i \tau
\tilde L_0 } = e^{ \, \xi L_X } e^{ \eta L_Y } e^{ \tau L_U } e^{ - i \tau 
\{ 2j \} } $. In the special cases $ j = \frac{3}{2}, \frac{4}{2},
\ldots $ this $ \tilde D_L^j $ gives the representation $D_L$, and for the
other values of $j$ $ \tilde D_L^j $ is a ray representation. The
infinitesimal generators of this ray representation, $ \tilde L_X = L_X $,
$ \tilde L_Y = L_Y $ and $ \tilde L_U = i \tilde L_0 $, form a central
extension of the Lie algebra $sl(2,\R)$: 
    \be
    [ \tilde L_U , \tilde L_X ] = 2 \tilde L_Y, \qquad
    [ \tilde L_X , \tilde L_Y ] = -2 \tilde L_U - 2 \{ 2j \} \, i, \qquad
    [ \tilde L_Y , \tilde L_U ] = 2 \tilde L_X,
    \l{cent}
the constant $ 2 \{ 2j \} $ plays the role of the commutator cocycle. (To
prove that $ \tilde D_L^j $ is indeed a ray representation it is enough to
verify that \r{cent} fulfils the conditions for a central extension, see
\cite{God}.) We can see that, by carrying out the shift $ \tilde L_X
\rightarrow L_X $, $ \tilde L_Y \rightarrow L_Y $, $ \tilde L_U
\rightarrow L_U $ on the central extension, we arrive at a representation
of the algebra $sl(2,\R)$. Conversely, starting from an arbitrary unitary
irreducible representation of $sl(2,\R)$ (indexed by a $ j > 1 $), an
appropriate redefinition yields such infinitesimal operators that can be
exponentiated to a continous unitary ray representation of the left
translation symmetry. In the cases $ j = \frac{3}{2}, \frac{4}{2}, \ldots
$ this ray representation is indeed a representation. 

The situation corresponds to the following general picture. A symmetry
group is represented on the Hilbert space of a quantum theory, in general,
by a continous unitary ray representation. The infinitesimal generators of
the ray representation give a central extension of the Lie algebra of the
group. If the symmetry group is semi-simple then the commutator cocycle
can be transformed out by an appropriate shift of the infinitesimal
generators, the new infinitesimal operators satisfy the commutation
relations of the Lie algebra with no central elements. Thus we arrive at a
unitary representation of the Lie algebra. 

Hence we can see that in \cite{JOT} only those cases are presented where a
representation of the symmetry group occurs, the ray representation cases
are not found. The reason of this is that in \cite{JOT} only those
irreducible unitary representations of $sl(2,\R)$ are considered, which
provide a representation of $SL(2,\R)$ as well. The analysis above shows
that the other $sl(2,\R)$ representations are also relevant in our
physical system. 
 
To make our considerations complete, in the end let us verify that the
phase factor appearing between $ \Psi_{m,E} (t = 2 \pi) $ and $ \Psi_{m,E}
(t = 0) $ in the ray representation cases does not lead to any physical
inconsistency. Fortunately, this phase factor is space-, $m$- and
$E$-independent, thus it is simply an overall constant phase factor for
all the wave functions. It causes a discontinuity of the wave functions
only in the time variable [at $ (t=2\pi) \equiv (t=0) $], while the inner
product involves integration over the space variables (cf. Sec.\ V), and
observables, such as the momentum or the angular momentum, include
derivations with respect to the space variables. A constant phase factor
does not change the expectation value of an observable or the inner
product of two wave functions, thus an expectation value or an inner
product calculated at $ t = 0 $ is the same as at $ t = 2 \pi $. So we can
see that the physically observable quantities allow the identification $
(t=2\pi) \equiv (t=0) $, the presence of the phase factor does not disturb
the physical interpretation issues and leads to no inconsistency. 

\bigskip

{\bf Acknowledgements:} The author wishes to thank L. Palla, Z. Horv\'ath, 
L. Szabados and G. Tak\'acs for valuable discussions.

\end{document}